\documentclass[prd,aps,showpacs,nofootinbib,preprint,eqsecnum]{revtex4}
\usepackage{graphicx,bm}
\usepackage{amsmath}
\usepackage{amssymb}
\usepackage{amsfonts}

\begin{document}

\def\pp{{\, \mid \hskip -1.5mm =}}
\def\cL{{\cal L}}
\def\be{\begin{equation}}
\def\ee{\end{equation}}
\def\bea{\begin{eqnarray}}
\def\eea{\end{eqnarray}}
\def\tr{\mathrm{tr}\, }
\def\nn{\nonumber \\}
\def\e{\mathrm{e}}

%%%
\def\Vec#1{\mbox{\boldmath $#1$}}
\def\Vecs#1{\mbox{\boldmath\tiny $#1$}}
\def\Lap{{\mathop{\Delta}\limits^{(3)}}}
%%%
%%%
\newcommand{\Eqn}[1]{&\hspace{-0.2em}#1\hspace{-0.2em}&}
\newcommand{\abs}[1]{\vert{#1}\vert}
%%%

%%%%%%%%%%%%%%%%%%%%%
%  Title
%%%%%%%%%%%%%%%%%%%%%
\title{Time-dependent matter instability and star singularity in $F(R)$
gravity
}

\author{Kazuharu Bamba$^{1,}$\footnote{
E-mail address: bamba@kmi.nagoya-u.ac.jp},
Shin'ichi Nojiri$^{1, 2,}$\footnote{E-mail address:
nojiri@phys.nagoya-u.ac.jp}
and Sergei D. Odintsov$^{3,}$\footnote{
E-mail address: odintsov@aliga.ieec.uab.es}$^{,}$\footnote{Also at Tomsk
State
Pedagogical University.}}
\affiliation{
$^1$Kobayashi-Maskawa Institute for the Origin of Particles and the
Universe,
Nagoya University, Nagoya 464-8602, Japan\\
$^2$Department of Physics, Nagoya University, Nagoya 464-8602, Japan\\
$^3$Instituci\`{o} Catalana de Recerca i Estudis Avan\c{c}ats (ICREA)
and Institut de Ciencies de l'Espai (IEEC-CSIC),
Campus UAB, Facultat de Ciencies, Torre C5-Par-2a pl, E-08193 Bellaterra
(Barcelona), Spain
}

%\date{\today}

%%%%%%%%%%%%%%%%%%%%%
%  Abstract
%%%%%%%%%%%%%%%%%%%%%
\begin{abstract}

We investigate a curvature singularity appearing in the star collapse 
process in $F(R)$ gravity. In particular, we propose 
an understanding of the 
mechanism to produce the curvature singularity. 
Moreover, we explicitly demonstrate that 
$R^\alpha$ ($1 < \alpha \leq 2$) term addition could cure the curvature 
singularity 
and viable $F(R)$ gravity models 
could become free of such a singularity. 
Furthermore, we discuss the realization process of the curvature singularity 
and estimate the time scale of its appearance. 
For exponential gravity, 
it is shown that in case of the star collapse, 
the time scale is much shorter than the age of the universe,
whereas in cosmological circumstances, 
it is as long as the cosmological time. 

\end{abstract}
%%%%%%%%%%%%%%%%%%%%%

%----------------------------
%\preprint{}
%----------------------------

\maketitle
%==============================================================================

%%%%%%%%%%%%%%%%%%%%%%%%%%%
%%%  Sec. I
%%%%%%%%%%%%%%%%%%%%%%%%%%%
\section{Introduction}

%%%%%
Modified gravity is expected to be the fundamental scenario to give the 
unified gravitational alternative to 
dark energy and inflation~\cite{Review-Nojiri-Odintsov}. A number of 
viable modified gravities are known. 
Of course, 
there exist 
important criteria for viability, 
such as the fulfillment of Solar System tests. 
Among these criteria, 
one of the most important ones is related with so-called 
matter instability~\cite{Dolgov:2003px} in $F(R)$ gravity, 
which means that the curvature inside matter sphere 
becomes very large, i.e., strong gravity 
(for an introduction to $F(R)$ gravity and very recent reviews on it, 
see, e.g.,~\cite{Review-Nojiri-Odintsov, Book-Capozziello-Faraoni}). 
It was indicated that such matter instability may be dangerous in the 
relativistic star formation process~\cite{maeda} due to the appearance of 
corresponding singularity. 
%%%%%
On the other hand, 
the hydrostatic equilibrium of a stellar structure in the framework of 
$F(R)$ gravity has been explored by studying the modified Lan\'{e}-Emden 
equation in Ref.~\cite{capozziello}. 

Recently, 
the instability in $F(R)$ gravity has been 
discussed for a gravitating system with a time dependent mass density 
like astronomical massive objects 
in Ref.~\cite{Arbuzova:2010iu}. 
It has been shown that 
when a star shrinks and the mass density becomes larger, the time-dependent 
matter instability, 
in which the scalar curvature becomes 
very large, could occur for a class of $F(R)$ gravity models. 
It is interesting to understand how common such time-dependent matter 
instability is in $F(R)$ gravity 
and how viable $F(R)$ gravity models may be protected against it. 
%%%%%
In this paper, we study the generation mechanism of 
the time-dependent matter instability in the star collapse. 
We show that the time-dependent matter instability develops 
and consequently the curvature singularity could appear 
in a viable 
$F(R)$ gravity model~\cite{Hu:2007nk} 
and some version of exponential gravity~\cite{Cognola:2007zu, 
Exponential-Gravity, Elizalde:2010ts}. 
%%%
We note that the equivalent or very similar modification of gravity 
to Ref.~\cite{Hu:2007nk} has been considered in 
Refs.~\cite{Starobinsky:2007hu, Appleby:2007vb}. 
%%%
In addition, we demonstrate that 
the curvature singularity could be cured by adding the 
higher derivative term $R^\alpha$ ($1 < \alpha \leq 2$) 
and viable $F(R)$ gravity models 
could become free of such a singularity. 
Furthermore, we discuss the realization process of the curvature singularity 
in the viable $F(R)$ gravity model 
and estimate the time scale of its appearance 
in exponential gravity. 
We show that in case of the star collapse, 
the time scale is much shorter than the age of the universe, 
whereas in cosmological circumstances, 
it is as long as the cosmological time. 
%%%%%
We mention that the problem of singularity in the star collapse 
was studied in Ref.~\cite{Frolov:2008uf}, 
in which it was stated that the curvature singularity in the future 
does not appear if the model of $F(R)$ gravity is built very carefully 
(for an earlier proposal, see~\cite{Briscese:2006xu}). 
It is considered that the curvature singularity could emerge in a generic 
$F(R)$ gravity model unless fine-tuning is taken. 
%%%%%
%%% Unit %%%
We use units of $k_\mathrm{B} = c = \hbar = 1$ and denote the
gravitational constant $8 \pi G$ by
${\kappa}^2 \equiv 8\pi/{M_{\mathrm{Pl}}}^2$
with the Planck mass of $M_{\mathrm{Pl}} = G^{-1/2} = 1.2 \times
10^{19}$GeV.
%%%%%%%%%%%%

%%%%%
The paper is organized as follows.
In Sec.\ II, we first review the matter instability and 
study the generation mechanism of 
the time-dependent matter instability. 
In Sec.\ III, we examine how the curvature singularity occurs 
and analyze the time scale of its appearance. 
Finally, conclusions are given in Sec.\ IV.
%%%%%

%%%%%%%%%%%%%%%%%%%%%%%%%%%
%%%  Sec. II
%%%%%%%%%%%%%%%%%%%%%%%%%%%
\section{Curvature singularity in the collapse of stars
}

%%%%%%%%%%%%%%%%%%%%%%%%%%%
%%%  Sec. II A
%%%%%%%%%%%%%%%%%%%%%%%%%%%
\subsection{Matter instability}

We start with reviewing the matter instability issue \cite{Dolgov:2003px} in 
$F(R)$ gravity. 
It is related with the fact that spherical body solution
in general relativity may not be the solution in modified theory. It may
appear when the energy density or the curvature is large compared with the
average one in the universe, as is the case inside of 
a star. 

We consider the following action
\be
\label{Dol1}
S = \int d^4 x \sqrt{-g}
\left\{ \frac{F(R)}{2\kappa^2} + \mathcal{L}_\mathrm{matter} \right\}\, ,
\ee
%%%%%
where $g$ is the determinant of the metric tensor $g_{\mu\nu}$, 
$F(R)$ is an arbitrary function of $R$, 
and 
$\mathcal{L}_\mathrm{matter}$ is the matter Lagrangian. 
%%%%%
The trace of 
the gravitational field equation derived from the action in Eq.~(\ref{Dol1}) 
is given by 
\be
\label{JGRG27}
\Box R + \frac{F^{(3)}(R)}{3 F^{(2)}(R)}\nabla_\rho R \nabla^\rho R
+ \frac{F'(R) R}{3F^{(2)}(R)} - \frac{2F(R)}{3 F^{(2)}(R)}
= \frac{\kappa^2}{3F^{(2)}(R)}T_\mathrm{matter}\, , 
\ee
where a prime denotes a derivative with respect to $R$,
${\nabla}_{\mu}$ is the covariant derivative operator associated with
$g_{\mu \nu}$, 
$\Box \equiv g^{\mu \nu} {\nabla}_{\mu} {\nabla}_{\nu}$
is the covariant d'Alembertian for a scalar field, 
and 
$T_\mathrm{matter} \equiv {T_\mathrm{matter}}_\rho^{\ \rho}$ 
is the trace of the matter energy-momentum 
tensor 
$T_{\mathrm{matter}\,\mu\nu}$ 
defined as 
$
T_{\mathrm{matter}\, \mu\nu} \equiv
-\left(2/\sqrt{-g}\right)
\left( \delta \mathcal{L}_\mathrm{matter} /\delta g^{\mu\nu} \right).
$
We also denote $d^nF(R)/dR^n$ by $F^{(n)}(R)$. 
Let us now 
examine the perturbation from the solution of general relativity.
We express
the scalar curvature solution given by the matter density in the
Einstein gravity by $R_{\mathrm{b}} \sim \kappa^2 \rho_{\mathrm{matter}}>0$
with $\rho_{\mathrm{matter}}$ being the energy density of matter
and
separate the scalar curvature $R$ into the sum of $R_{\mathrm{b}}$
and the perturbed part $R_{\mathrm{p}}$ as $R=R_{\mathrm{b}} + R_{\mathrm{p}}$ 
$\left(\left|R_{\mathrm{p}}\right|\ll
\left|R_{\mathrm{b}} \right|\right)$. 
Substituting this expression into 
Eq.~(\ref{JGRG27}), 
we obtain the perturbed
equation~\cite{Nojiri:2003ft, Nojiri:2007as}. 
It is convenient to consider the case that $R_{\mathrm{b}}$ and 
$R_{\mathrm{p}}$ are homogeneous and 
isotropic, that is, they do not depend on the spatial coordinates.
Hence, the d'Alembertian can be replaced with the
second derivative with respect to the time coordinate:
$\Box R_{\mathrm{p}} \to - \partial_t^2 R_{\mathrm{p}}$. 
As a result, the perturbed equation 
has the following structure: 
\be
\label{JGRG30}
0=-\partial_t^2 R_{\mathrm{p}} + \mathcal{U}(R_{\mathrm{b}}) R_{\mathrm{p}} 
+ \mathrm{const.}\, .
\ee
Thus, 
if $\mathcal{U}(R_{\mathrm{b}})>0$, $R_{\mathrm{p}}$ becomes exponentially 
large with time $t$: 
$R_{\mathrm{p}} \sim \e^{\sqrt{\mathcal{U}(R_{\mathrm{b}})} t}$ 
and the system is unstable. 

For example,
in the model of
$F(R) = R-a/R + bR^2$ with $a$ and $b$ being positive
constants~\cite{Nojiri:2003ft},
for $b \gg a/\abs{ R_{\mathrm{b}}^3 }$ one gets
\be
\label{JGRG33}
\mathcal{U}(R_{\mathrm{b}})\sim \frac{R_{\mathrm{b}}}{3}>0\, .
\ee 
Therefore,
the system could be unstable. 
Since $R_{\mathrm{b}}$ is estimated as~\cite{Dolgov:2003px}, however,
\be
\label{JGRG32}
R_{\mathrm{b}} \sim \left(10^3 \mbox{sec}\right)^{-2}
\left(\frac{\rho_\mathrm{matter}}{\mbox{g\,cm}^{-3}}\right)\, ,
\ee 
the decay time is $\sim$ $1,000$ sec, 
i.e., 
macroscopic.
On the other hand, in a viable model~\cite{Hu:2007nk}
\be
F(R) = R-m^2 \frac{c_1 \left(R/m^2 \right)^n}{c_2
\left(R/m^2 \right)^n +1}\,,
\label{eq:H-S}
\ee
where $c_1$ and $c_2$ are dimensionless parameters,
$n ( > 0 )$ is a positive constant, and
$m$ is a mass scale,
if one takes $c_1 >0$,
$\mathcal{U}(R_{\mathrm{b}})$ is negative as
\be
\label{JGRG34}
\mathcal{U}(R_{\mathrm{b}}) \sim - \frac{(n+2) c_2^2 m^2}{n(n+1) c_1} < 0 \, .
\ee 
Consequently,
the system could be stable and there is no matter instability.

We remark that $-\mathcal{U}(R_{\mathrm{b}})$ can be regarded as the square of 
the effective mass 
$m_\mathrm{eff}^2$
for the scalar mode $R_{\mathrm{p}}$,
given by
$m_\mathrm{eff}^2 \approx
F^{\prime \prime}(R)/3 
$~\cite{M-I-F-S}. 
This means that
if 
$F^{\prime \prime}(R) >0$, 
one has $m_\mathrm{eff}^2 > 0$ and hence
the scalar mode $R_{\mathrm{p}}$ is stable,
whereas it is unstable for
$F^{\prime \prime}(R) <0$ 
because
$m_\mathrm{eff}^2 < 0$.
If 
$F^{\prime \prime}(R) <0$, 
this instability occurs inside matter when the deviation of $R$ from
$R_{\mathrm{b}} \sim \kappa^2 \rho_\mathrm{matter}$ emerges.
Thus, it can be interpreted as a matter instability.
%%%%%

%%%%%%%%%%%%%%%%%%%%%%%%%%%
%%%  Sec. II B
%%%%%%%%%%%%%%%%%%%%%%%%%%%
\subsection{Time-dependent matter instability and curvature singularity
}

In Ref.~\cite{Arbuzova:2010iu}, it has been shown that even for 
the viable models such as those in Refs.~\cite{Hu:2007nk, Starobinsky:2007hu}, 
the increasing matter density could generate the curvature singularity,
which we will describe in this subsection. 
We write $F(R)$ as\footnote{
Note that the definition of $F(R)$ and $f(R)$ are different
from those in Ref.~\cite{Arbuzova:2010iu}.}
\be
\label{Dol2}
F(R) = R + f(R)\, .
\ee 
The trace equation of the gravitational field is described as 
\be
\label{Dol3} 
3 \Box f'(R) - R + Rf'(R) - 2f(R) = \kappa^2 T_{\mathrm{matter}}\, .
\ee
We now consider a small region inside the star, where the energy density
can be regarded as 
homogeneous and isotropic as assumed in Ref.~\cite{Dolgov:2003px}. 
%%%%%
Since we study the gravitational field of the star such as 
the Sun, whose mass density is 
$\rho_{\odot} = 1.4 \,\mathrm{g}\,\mathrm{cm}^{-3}$, 
we also suppose that the gravitational field is weak 
and the curvature is small because 
$R_{\mathrm{b}} \sim \kappa^2 \rho_\mathrm{matter}$. 
(If a neutron star has the solar mass $M_{\odot} = 2.0 \times 10^{33} 
\,\mathrm{g}$ and the radius 
$1.0 \times 10^6 \mathrm{cm}$, the mass density is 
$4.7 \times 10^{14} \,\mathrm{g}\,\mathrm{cm}^{-3}$, 
which is much larger than the solar mass density. 
The gravitational field of a neutron star is regarded as strong because 
a general relativistic treatment is necessary to explore a neutron star.)
%%%%%
Then, we may replace $\Box$ in Eq.~(\ref{Dol3}) with $- \partial_t^2$.
By defining $\varphi$ 
as $\varphi\equiv - f'(R)$, we can (at least locally) solve $R$
with respect to $\varphi$: $R=R(\varphi)$. 
Using this solution,
Eq.~(\ref{Dol3}) can be rewritten as
\be
\label{Dol4}
\ddot \varphi = \mathcal{F}(\varphi,t) \equiv \frac{1}{3}
\left\{ R\left(\varphi\right) + R\left(\varphi\right) \varphi
+ 2 f \left( R\left(\varphi\right) \right) + \kappa^2 T_{\mathrm{matter}}(t)
\right\}\, , 
\ee
where the dot denotes the time derivative of $\partial/\partial t$. 
Compared with Newton's equation of motion $m\ddot{\bm{x}} = \bm{F}$, we can
regard $\mathcal{F}(\varphi,t)$
as a ``force''. From Eq.~(\ref{Dol4}), we also obtain
\be
\label{Dol5}
0 = \frac{1}{2} {\dot\varphi}^2 - \int dt \dot\varphi(t)
\mathcal{F}(\varphi(t),t) \, .
\ee
We explore the case in which
$T_{\mathrm{matter}}$ increases with time:
\be
\label{Dol5b}
T_{\mathrm{matter}} = - T_{\mathrm{matter}\,0} \left( 1 + \frac{t}{t_0}
\right)\, , 
\ee
where $T_{\mathrm{matter}\,0}$ is constant and 
$t_0$ is a time.

%%%%%%%%%%%%%%%%%%%%%%%%%%%
%%%  Sec. II B 1
%%%%%%%%%%%%%%%%%%%%%%%%%%%
%\subsubsection{Viable model}

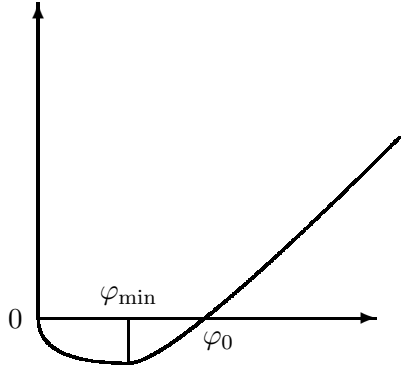
\begin{figure}[ht]
\begin{center}

\unitlength=0.6mm
\begin{picture}(100,100)

\thicklines
\put(10,15){\vector(1,0){75}}
\put(10,15){\vector(0,1){70}}
\put(5,15){\makebox(0,0){$0$}}
\qbezier(10,15)(10,5)(30,5)
\qbezier(30,5)(40,5)(90,55)
\put(30,20){\makebox(0,0){$\varphi_\mathrm{min}$}}
\put(50,10){\makebox(0,0){$\varphi_0$}}

\thinlines
\put(30,5){\line(0,1){10}}

\end{picture}

\caption{The ``potential'' $U\left(\varphi\right)$ in Eq.~(\ref{Dol10}).
The vertical and horizontal axes show
$U\left(\varphi\right)$ and $\varphi$, respectively.
\label{Fig1}}
\end{center}
\end{figure}

In case of the viable models, 
e.g., those in Refs.~\cite{Hu:2007nk, Starobinsky:2007hu}, 
$f(R)$ is given by
\be
\label{Dol6}
f(R) \sim - f_0 + \frac{f_1}{n R^n} \, ,
\ee
for large $R$. 
%%%%%
Here, $f_0$ and $f_1$ are constants. 
%%%%%
If $f(R) = F(R) -R$ is written by Eq.~(\ref{eq:H-S}), one finds
$f_0 = \left(c_1/c_2 \right)m^2$ and
$f_1/n = \left(c_1/c_2^2 \right)m^{2\left(n+1\right)}$.
%%%%%
In this case, we have
\be
\label{Dol7}
\varphi = \frac{f_1}{R^{n+1}}\, ,
\ee
and therefore $\mathcal{F}(\varphi,t)$ in Eq.~(\ref{Dol4}) is given by
\be
\label{Dol8}
\mathcal{F}(\varphi,t) = \frac{1}{3} \left\{ \left( \frac{\varphi}{f_1}
\right)^{-\frac{1}{n+1}}
+ f_1 \left( \frac{\varphi}{f_1} \right)^{\frac{n}{n+1}} - 2 f_0
+ \frac{2 f_1}{n} \left( \frac{\varphi}{f_1} \right)^{\frac{n}{n+1}}
 - \kappa^2 T_{\mathrm{matter}\,0} \left( 1 + \frac{t}{t_0} \right)
\right\}\, .
\ee
Combining Eqs.~(\ref{Dol5}) and (\ref{Dol8}), we obtain 
\bea
\label{Dol9}
\mathcal{E} &=& \frac{1}{2} {\dot\varphi}^2 + U\left(\varphi,t\right)\, ,\nn
U\left(\varphi,t\right) &=& \frac{1}{3} \left\{ - \frac{(n+1)f_1}{n}\left(
\frac{\varphi}{f_1} \right)^{\frac{n}{n+1}}
 - \frac{f_1^2(n+1)}{2n + 1} \left( 1 + \frac{2}{n} \right)
\left( \frac{\varphi}{f_1} \right)^{\frac{2n + 1}{n+1}} \right. \nn
&& \left. + \left[ \kappa^2 T_{\mathrm{matter}\,0}\left( 1 + \frac{t}{t_0}
\right) + 2 f_0 \right] \varphi
 - \frac{\kappa^2 T_{\mathrm{matter}\,0}}{t_0} \int dt \varphi(t)
\right\}\, ,
\eea 
where
$\mathcal{E}$ is a constant corresponding to ``energy'' in the classical
mechanics
and $U(\varphi)$ to the potential.
The case that $R$ is large corresponds to the case that $\varphi$ is small. 
%%%%%
Hence, 
we neglect the second term on the right-hand side of the expression for 
$U\left(\varphi,t\right)$ in (\ref{Dol9}), 
i.e., 
$-\left(1/3\right) \left[ f_1^2(n+1)/\left( 2n + 1 \right) \right] 
\left( 1 + 2/n \right) \left( \varphi/f_1 
\right)^{\left(2n + 1\right)/\left(n+1\right)}$. 
%%%%%
First, we consider the case that $T_{\mathrm{matter}\,0}=0$. 
It follows from Eq.~(\ref{Dol9}) that
the ``potential'' $U(\varphi)$ is given by
\be
\label{Dol10}
U\left(\varphi\right) = \frac{1}{3} \left\{ - \frac{n+1}{n}\left(
\frac{\varphi}{f_1} \right)^{\frac{n}{n+1}}
+ 2 f_0 \varphi \right\}\, ,
\ee
which vanishes at
\be
\label{Dol11}
\varphi=0\, ,\quad \varphi_0 \equiv
f_1^{-n}
\left( \frac{n+1}{2 n f_0} \right)^{n+1} \, ,
\ee
and has a minimum at
\be
\label{Dol12}
\varphi = \varphi_\mathrm{min} \equiv f_1^{-n}
\left( 2 f_0 \right)^{-(n+1)}\, .
\ee
The conceptual form of $U\left(\varphi\right)$ is 
shown in Fig.~\ref{Fig1}. 
It is clear from Fig.~\ref{Fig1} that 
if we start with an initial condition, for example, $\varphi>\varphi_0$ and
$\dot\varphi \leq 0$,
$\varphi$ reaches $\varphi=0$, which corresponds to the curvature
singularity $R=\infty$.

\begin{figure}[ht]
\begin{center}

\unitlength=0.6mm
\begin{picture}(100,100)

\thicklines

\put(10,15){\vector(1,0){75}}
\put(10,15){\vector(0,1){70}}
\put(5,15){\makebox(0,0){$0$}}
\qbezier(10,15)(10,5)(30,5)
\qbezier(30,5)(40,5)(90,55)

\qbezier(10,15)(10,8)(25,8)
\qbezier(25,8)(30,8)(90,68)

\qbezier(10,15)(10,13)(20,13)
\qbezier(20,13)(25,13)(90,78)

\thinlines

\put(88,55){\vector(0,1){10}}
\put(88,68){\vector(0,1){6}}

\end{picture}

\caption{The change of the ``potential'' $U\left(\varphi\right)$
in Eq.~(\ref{Dol9}) with time $t$.
Legend is the same as Fig.~\ref{Fig1}.
\label{Fig2}}
\end{center}
\end{figure}
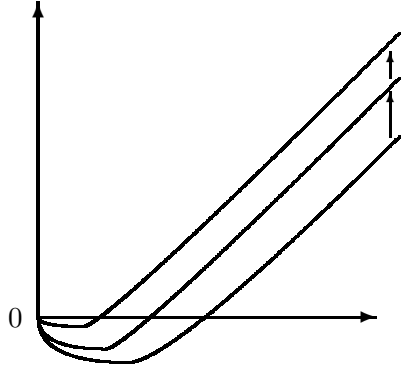

When $T_{\mathrm{matter}\,0}\neq 0$, the potential shifts up with time as in
Fig.~\ref{Fig2}.
By growing the energy density, the singularity $R=\infty$ can be realized
easily. 
Even if $T_{\mathrm{matter}}=0$, there could appear the curvature
singularity but the increase of $T_{\mathrm{matter}}$,
which may occur by the star collapse, 
becomes a trigger to generate
the singularity.

%%%%%%%%%%%%%%%%%%%%%%%%%%%
%%%  Sec. II B 2
%%%%%%%%%%%%%%%%%%%%%%%%%%%
%\subsubsection{$R^\alpha$ term for the avoidance of the curvature
%singularity}

A prescription to avoid the singularity is to grow 
the potential at $R=\infty$.
In order to realize the growing up of the potential, we add the following
term, say,
to $f(R)$ in Eq.~(\ref{Dol6})
%.
\be
\label{Dol13}
f(R) \to f(R) - f_2 R^\alpha \, .
\ee
Here, we assume $f_2$ is a positive constant: $f_2>0$ 
and $\alpha$ is a constant. 
If we choose $\alpha>1$, the added term dominates when the curvature $R$ is
large and we find
\be
\label{Dol14}
\varphi \sim \alpha f_2 R^{\alpha -1} \, .
\ee 
Hence,
the case $R\to \infty$ corresponds to the case $\varphi\to +\infty$.
In this case, $\mathcal{F}(\varphi,t)$ in Eq.~(\ref{Dol4}) is given by
\be
\label{Dol15}
\mathcal{F}(\varphi,t) \sim f_2 \left( \alpha - 2 \right)
\left(\frac{\varphi}{\alpha f_2}\right)^{\frac{\alpha}{\alpha - 1}} \, .
\ee
Since $\alpha>0$, the magnitude of the ``force'' $\mathcal{F}(\varphi,t)$
becomes infinitely large
when $R\to + \infty$.
When $\alpha<2$, the force is negative and works to decrease $\varphi$ or
the scalar curvature, and
therefore the curvature singularity is not realized. 
In fact, from Eq.~(\ref{Dol5}) we have the following form: 
\be
\label{Dol15b}
\mathcal{E} = \frac{1}{2} {\dot\varphi}^2 + U\left(\varphi,t\right)\, ,\quad
U\left(\varphi\right) \sim
\frac{f_2^2 \alpha \left(\alpha - 1\right) \left( 2 - \alpha \right)}{2
\alpha - 1}
\left(\frac{\varphi}{\alpha f_2}\right)^{\frac{2 \alpha - 1}{\alpha - 1}}\,
.
\ee 
Thus, 
the ``potential'' $U(\varphi)$ is positive as long as $1<\alpha<2$ and
becomes infinite
when $\varphi \to \infty$, which 
means that we need infinite ``work'' to arrive at
$\varphi \to \infty$.
On the other hand, when $\alpha>2$, the force is positive and therefore the
force makes the curvature
infinite and the curvature singularity is easily realized.
Note that even if we choose $f_2$ to be negative, the result is not changed 
because the curvature 
singularity corresponds to $\varphi\to - \infty$ in this case although the 
sign of the ``force'' 
$\mathcal{F}(\varphi,t)$ is changed. 

%%%%%
It is also important to remark the case of adding $R^2$ term, i.e.,
$\alpha = 2$ as
$f(R) \to f(R) + f_3 R^2$, where $f_3$ is assumed to be a positive constant.
For the ``realistic'' model in Eq.~(\ref{Dol6}),
since the inverse power term $f_1/\left(n R^n\right)$
can be negligible when the curvature is large,
we may consider the model $F(R) \sim R + f_3 R^2$.
In this case, we may regard $\varphi \sim - 2 f_3 R$.
In the expression of the ``force'' $\mathcal{F}(\varphi,t)$
in Eq.~(\ref{Dol4}), the second term $R\left(\varphi\right) \varphi$ is
cancelled by the third term $2 f \left( R\left(\varphi\right) \right)$,
which is consistent with Eq.~(\ref{Dol15}), and the fourth term
$\kappa^2 T_{\mathrm{matter}}(t)$ could be neglected.
Hence,
only the first term $R\left(\varphi\right)$ contributes to the dynamics and
it gives the ``force'' linear to $\varphi$ as
$\mathcal{F}(\varphi,t) \sim -\varphi/\left(6 f_3 \right)$.
Thus, from Eq.~(\ref{Dol5})
we obtain quadratic ``potential''
$U(\varphi) \sim \varphi^2/\left(12 f_3 \right)$,
which becomes positive infinity when $R$ goes to positive infinity
($\varphi$ goes to negative infinity).
As a result, the curvature singularity is prevented.
%%%%%

%%%%%%%%%%%%%%%%%%%%%%%%%%%
%%%  Sec. II B 3
%%%%%%%%%%%%%%%%%%%%%%%%%%%
%\subsubsection{Exponential gravity  model}

Instead of the model in Eq.~(\ref{Dol6}), we may 
investigate the exponential gravity 
model~\cite{Cognola:2007zu, Exponential-Gravity, Elizalde:2010ts},
\be
\label{Dol16}
f(R) = -f_{\mathrm{c}} \left( 1 - \e^{-\frac{R}{R_{\mathrm{c}}}} \right)\, ,
\ee 
where
$f_{\mathrm{c}}$ and $R_{\mathrm{c}}$ are positive constants.
In the model in Eq.~(\ref{Dol16}), we have
\be
\label{Dol17}
\varphi = \frac{f_{\mathrm{c}}}{R_{\mathrm{c}}} 
\e^{-\frac{R}{R_{\mathrm{c}}}}\, .
\ee
Therefore, the curvature singularity $R\to +\infty$ corresponds to
$\varphi=+ 0$.
For the model in Eq.~(\ref{Dol16}), we find
\be
\label{Dol18}
\mathcal{F}(\varphi,t) \sim - \frac{R_{\mathrm{c}}}{3} 
\ln \frac{R_{\mathrm{c}} \varphi}{f_{\mathrm{c}}} \, .
\ee
When $\varphi\to +0$, the ``force'' $\mathcal{F}(\varphi,t)$ becomes
positive and infinite, 
but now it follows from Eq.~(\ref{Dol5})
that we have the following form:
\be
\label{Dol19}
\mathcal{E} = \frac{1}{2} {\dot\varphi}^2 + U\left(\varphi,t\right)\, ,\quad
U\left(\varphi\right) \sim \frac{R_{\mathrm{c}}}{3}\varphi \left( 
\ln \frac{R_{\mathrm{c}}\varphi}{f_{\mathrm{c}}} - 1 \right)\, ,
\ee
which gives the finite value of ``potential'' when $\varphi\to +0$. 
This implies
that we only need finite ``work'' to arrive at $\varphi=+0$.
Therefore, the curvature singularity $R\to \infty$ can be realized.
This situation is almost the same as that 
in the model in Eq.~(\ref{Dol6}) 
because when $\varphi\to +0$, 
the ``force'' in 
Eq.~(\ref{Dol8}) diverges but the ``potential'' in Eq.~(\ref{Dol9}) is 
finite.

%%%%%
We mention that there exists
the following model
with two exponential terms to realize inflation as well as
the late-time cosmic acceleration~\cite{Elizalde:2010ts}
\be
f(R) = -f_{\mathrm{c}} \left( 1 - \e^{-\frac{R}{R_{\mathrm{c}}}} \right)
-f_\mathrm{i} \left[ 1 - \e^{-\left(\frac{R}{R_\mathrm{i}}\right)^q}
\right] + \gamma R^\beta\, ,
\label{eq:Dol-B-1}
\ee
where $f_\mathrm{i}$, $R_\mathrm{i} (\gg R_{\mathrm{c}})$ and $\gamma$ are 
positive constants, $\beta$ is a constant, and $q (>1)$ is a natural number.
Here, we take $R_{\mathrm{c}}$ and $R_\mathrm{i}$ as the current curvature and
the value of $R$ during inflation.
If (i) $R \lesssim R_{\mathrm{c}}$,
(ii) $R_{\mathrm{c}} \ll R \ll R_\mathrm{i}$,
and
(iii) $R \gg R_\mathrm{i}$,
the first, second, and third terms
on the right-hand side of Eq.~(\ref{eq:Dol-B-1}) becomes dominant,
respectively.
For the case (i), the model in Eq.~(\ref{eq:Dol-B-1})
corresponds to the one in Eq.~(\ref{Dol16}),
whereas for the case (iii)
the behavior of the model in Eq.~(\ref{eq:Dol-B-1})
is similar to that in the case of Eq.~(\ref{Dol13}) with
the additional term domination.
In terms of the case (ii),
we examine the behavior for the corresponding form of $f(R)$ as
$f(R) =-f_\mathrm{i} \left[ 1 - \e^{-\left(R/R_\mathrm{i}\right)^q}
\right]$.
In this case, we obtain 
\be
\varphi 
\sim \frac{q f_\mathrm{i}}{R_\mathrm{i}}
\left(\frac{R}{R_\mathrm{i}}\right)^{q-1} \, ,
\label{eq:Dol-B-2}
\ee
where we have used $R \ll R_\mathrm{i}$. 
For $q>1$, $\varphi$ increases as $R$ becomes large.
Moreover, we find 
\be
\mathcal{F}(\varphi,t) \sim - \frac{q f_\mathrm{i}}{3}
\left(\frac{R_\mathrm{i}\varphi}{q
f_\mathrm{i}}\right)^{2q/\left(q-1\right)}
\, .
\label{eq:Dol-B-3}
\ee
The sign of the ``force'' $\mathcal{F}(\varphi,t)$ is negative
and therefore the force works to decrease $\varphi$. 
Furthermore, 
Eq.~(\ref{Dol5}) becomes 
\be
\mathcal{E} = \frac{1}{2} {\dot\varphi}^2 + U\left(\varphi,t\right)\, ,\quad
U\left(\varphi\right) \sim \frac{q-1}{3\left(3q-1\right)}
\frac{\left(q f_\mathrm{i}\right)^2}{R_\mathrm{i}}
\left( \frac{R_\mathrm{i} \varphi}{q f_\mathrm{i}}
\right)^{\left(3q-1\right)/\left(q-1\right)}
\, .
\label{eq:Dol-B-4}
\ee
Since $q>1$, $U\left(\varphi\right)$ increases as $\varphi$ becomes large.
Consequently, the behavior of the model in Eq.~(\ref{eq:Dol-B-1})
in the intermediate regime $R_{\mathrm{c}} \ll R \ll R_\mathrm{i}$
is stable.
%%%%%
Note that this does not mean that the curvature singularity could not be 
generated because we are assuming 
the scalar curvature is finite as $R \ll R_\mathrm{i}$ although the 
curvature has a tendency 
to decrease by the ``force'' $\mathcal{F}(\varphi,t)$. Without the last term
in Eq.~(\ref{eq:Dol-B-1}),
there could appear the curvature singularity when $R\gg R_\mathrm{i}$.

%%%%%
In Ref.~\cite{Arbuzova:2010iu}, 
the emergence of the curvature singularity in the future 
has been investigated by the analytical studies, 
which have also been supplemented with numerical solutions. 
The equation for the evolution of $R$ was reduced to 
the form of an oscillator equation $\ddot{x} + dV(x)/dx = 0$, 
where $x$ and $V$ corresponds to $\varphi$ and $U$ in this paper. 
In case of a time dependent potential $V = V(x,t)$, 
since it is impossible to obtain an analytical solution, 
the qualitative behavior of the solution has been analyzed. 
%%%%%

%%%%%%%%%%%%%%%%%%%
%%%  Sec. III
%%%%%%%%%%%%%%%%%%%
\section{Realization of the curvature singularity}

%%%%%%%%%%%%%%%%%%%
%%%  Sec. III A
%%%%%%%%%%%%%%%%%%%
\subsection{Realization process of the curvature singularity}

We discuss how the curvature singularity is realized.
%%%%%
As an example, we investigate 
the viable model in Eq.~(\ref{Dol6}).
%%%%%
We consider a small region inside 
the star, which can be regarded as 
homogeneous and isotropic. 
In this case, 
the space-time is locally described by 
the flat Friedmann-Lema\^{i}tre-Robertson-Walker (FLRW): 
\be
\label{Dol20}
ds^2 = - dt^2 + a^2(t) \sum_{i=1,2,3} \left(dx^i\right)^2\, , 
\ee
where $a(t)$ is the scale factor. 
Note that since we are examining the star collapse, 
the Hubble rate $H\equiv \dot a/ a$
is negative, that is, the space-time is shrinking.
We also mention 
that the energy densities of the matters automatically increase 
because the region is shrinking. 
Thus, 
we do not give an explicit time dependence as in Eq.~(\ref{Dol5b}).

In case of 
cosmology, it is well-known that for the model in Eq.~(\ref{Dol6}),
Type II singularity can be realized, where the scale factor $a$ 
and the effective energy density are finite but the effective pressure
and the scalar curvature $R$ diverge~\cite{Singularity-F(R)}. 
%%%%%
We apply this fact to the star collapse. 
%%%%%
Since $a$ is finite, the energy density and the pressure from the matter are
finite and can be
neglected near the singularity. 
In such a situation,
we find the Hubble rate
$H$ is given by
\be
\label{Dol21}
H \sim - \frac{h_{\mathrm{st}}}{ \left( t_{\mathrm{st}} - t 
\right)^{\frac{n}{n+2}}}\, ,
\ee 
where
$h_{\mathrm{st}}$ is a positive constant and 
$t_{\mathrm{st}}$ is the time when the curvature 
singularity in the star appears. 
When $t\to t_{\mathrm{st}}$, $H$ is finite but $\dot H$ and therefore
the scalar curvature $R=6\dot H + 12 H^2$ diverge. 
This implies 
that in the model in Eq.~(\ref{Dol6}), the naked curvature singularity
is generated in the finite future.
In the above analysis, we have assumed that the region we are considering is
almost homogeneous and isotropic. If the assumptions are valid near the
singularity, the singularity appears simultaneously
anywhere in the region. Even if the assumptions of the homogeneity 
and isotropy are broken, the
naked singularities appear densely in the region.
Furthermore, 
we remark 
that in general, the matter density becomes larger in the region
deeper from the surface of the star. 
Hence, 
we expect that first, the naked curvature singularity
could be generated near the center of the star. If the singularity generates
the attractive
force, the star shrinks more, but if the generated force is repulsive, there
might occur the
explosion. If the explosion could occur, however, $H$ must change its sign 
from negative to positive, 
which seems to be difficult to be realized.

On the other hand, 
the model in Eq.~(\ref{Dol6}) could generate the cosmological 
singularity, 
where the Hubble rate behaves similar to Eq.~(\ref{Dol21}) as
\be
\label{Dol22}
H \sim \frac{h_{\mathrm{co}}}{\left( t_{\mathrm{co}} - t 
\right)^{\frac{n}{n+2}}}\, .
\ee
In the expanding cosmology, $H$ should be positive 
and hence 
the constant $h_{\mathrm{co}}$ is positive. 
Moreover, the singularity in the universe occurs at 
$t=t_{\mathrm{co}}$. 
The values of $t_{\mathrm{co}}$ (and $h_{\mathrm{co}}$) in Eq.~(\ref{Dol22})
and $t_{\mathrm{st}}$ (and $h_{\mathrm{st}}$) in Eq.~(\ref{Dol21})
could be determined
by the initial condition for $H$ and $\dot H$. Since we are considering the
collapsing star,
the absolute values of $H$ and $\dot H$ inside the star are expected to be
large compared
with the values in the expanding universe. Therefore, we expect 
$t_{\mathrm{co}}>t_{\mathrm{st}}$,
that is, the curvature singularity in the star occurs before the cosmological
singularity. 

%%%%%
We note that
in order to remove the finite-time future singularity,
adding a $R^2$ term to the form of $f(R)$ works 
[the first reference in 
Ref.~\cite{Review-Nojiri-Odintsov},~\cite{Singularity-F(R)}]. 
Similarly,
the $R^2$ term can cure
the curvature singularity under consideration in this paper,
as analyzed in Sec. II B.
The $R^2$ term shifts the value of the potential near the curvature singularity
in the scalar tensor description.
This fact seems to show a close relation between
the curvature singularity in the star under discussion
and the finite-time future singularity in the context of
cosmology.
%%%%%

%%%%%%%%%%%%%%%%%%%
%%%  Sec. III B
%%%%%%%%%%%%%%%%%%%
\subsection{Time scale for the realization of the curvature singularity}

%%%%%
We demonstrate the estimation of the time scale for the realization 
of the curvature singularity by exploring the exponential gravity model 
in Eq.~(\ref{Dol16}). 
%%%%%
In the flat FLRW background~(\ref{Dol20}), 
the gravitational field equations derived from the action in Eq.~(\ref{Dol1}) 
read 
\bea
0 \Eqn{=} -\frac{1}{2}F(R) +3\left(H^2 + \dot{H} \right) F^{\prime}(R)
-18\left(4 H^2 \dot{H} + H \ddot{H} \right) F^{\prime \prime}(R)
+ \kappa^2 \rho_{\mathrm{matter}}\,,
\label{eq:B-3B-1} \\
0 \Eqn{=} \frac{1}{2}F(R) - \left(\dot{H} +3H^2 \right) F^{\prime}(R)
+6\left(8 H^2 \dot{H} +6H\ddot{H} +4\dot{H}^2 + \dddot{H} \right)
F^{\prime \prime}(R)
\nonumber \\
&&
{}+36\left(4 H \dot{H} + \ddot{H} \right)^2
F^{\prime \prime \prime}(R)
+ \kappa^2 P_{\mathrm{matter}}\,,
\label{eq:B-3B-2}
\eea
where $P_{\mathrm{matter}}$ is the pressure of matter.
Substituting Eq.~(\ref{Dol2}) with Eq.~(\ref{Dol16}) into
Eq.~(\ref{eq:B-3B-1}), we find
\be
0 = -3H^2 + f_{\mathrm{c}} \left\{
\frac{1}{2} + \left[ -\frac{1}{2}-3\left(H^2 + \dot{H} \right) 
\frac{1}{R_{\mathrm{c}}}
-18\left(4 H^2 \dot{H} + H \ddot{H} \right) \frac{1}{R_{\mathrm{c}}^2}
\right] \e^{-\frac{R}{R_{\mathrm{c}}}}
\right\} + \kappa^2 \rho_{\mathrm{matter}}\,.
\label{eq:B-3B-3}
\ee
We examine the case $R\to \infty$.
In this case, it follows from Eq.~(\ref{eq:B-3B-3}) that
$H$ is expressed as
\be
H \sim A \left(t_{\mathrm{s}} - t \right) \ln \frac{t_{\mathrm{s}} - t}{t_1}
+ H_{\mathrm{s}}\,,
\label{eq:B-3B-4}
\ee
where $A$ is a constant, $t_{\mathrm{s}}$ is the time when the curvature
singularity appears, $t_1$ is a time,
and $H_{\mathrm{s}}$ is the value of $H$ at
$t=t_{\mathrm{s}}$.
When $R\to \infty$, $t \to t_{\mathrm{s}}$ and $H \to H_{\mathrm{s}}$.
In the limit of $t \to t_{\mathrm{s}}$,
by combining Eq.~(\ref{eq:B-3B-4}) with Eq.~(\ref{eq:B-3B-3}) , we obtain
\be
0 = -3H_{\mathrm{s}}^2 + f_{\mathrm{c}} \left(
\frac{1}{2} -\frac{3H_{\mathrm{s}}}{R_{\mathrm{c}} t_1}
\e^{-\frac{12H_{\mathrm{s}}^2}{R_{\mathrm{c}}}+1}
\right) + \kappa^2 \rho_{\mathrm{matter}} +
O\left(t_{\mathrm{s}} - t \right)\,,
\label{eq:B-3B-5}
\ee
where we have taken $A=R_{\mathrm{c}}/6$ and 
used the fact that 
the term proportional to $\ddot{H}$ is dominant over
other terms in Eq.~(\ref{eq:B-3B-3})
because $\ddot{H} \sim A/\left(t_{\mathrm{s}} - t \right)$ and
$\dot{H} \sim -A\left\{
\ln \left[ \left(t_{\mathrm{s}} - t \right)/t_1 \right] +1
\right\}$.
As an initial condition, we set the initial time $t=0$ and
$\dot{H} =0$ at $t=0$. 
By deriving the solution of 
$\dot{H} =0$ at $t=0$ in terms of $t_1$, we have 
$t_1 = e t_{\mathrm{s}}$. 
Moreover, by solving Eq.~(\ref{eq:B-3B-5}) in terms of $t_1$,
we acquire
\be
t_1 = \frac{6H_{\mathrm{s}}}{R_{\mathrm{c}}} \mathcal{I}
\e^{-\frac{12H_{\mathrm{s}}^2}{R_{\mathrm{c}}}+1}\,,
\label{eq:B-3B-6}
\ee
where
\be
\mathcal{I} \equiv
\frac{f_{\mathrm{c}}}{f_{\mathrm{c}} -6H_{\mathrm{s}}^2
+2\kappa^2 \rho_{\mathrm{matter}}}\,.
\label{eq:B-3B-7}
\ee
If we regard $R_{\mathrm{c}}$ as the current curvature, we get
$R_{\mathrm{c}} \approx 12 H_{\mathrm{c}}^2$, where
$H_{\mathrm{c}} = 2.1 h \times 10^{-42} \mathrm{GeV}$~\cite{Kolb and Turner}
with $h = 0.7$~\cite{Freedman:2000cf, Komatsu:2010fb}
is the current value of the Hubble rate. 
{}From $t_1 = e t_{\mathrm{s}}$ and Eq.~(\ref{eq:B-3B-6}),
we find
\be
\frac{t_{\mathrm{s}}}{t_{\mathrm{a}}} =
\frac{1}{2}
\frac{H_{\mathrm{c}}^{-1}}{t_{\mathrm{a}}}
\frac{H_{\mathrm{s}}}{H_{\mathrm{c}}}
\e^{-\left( \frac{H_{\mathrm{s}}}{H_{\mathrm{c}}} \right)^2} \mathcal{I}\,,
\label{eq:B-3B-8}
\ee
where
$t_{\mathrm{a}} = 
2.9 \times 10^{17}
\,\mathrm{sec}$~\cite{Kolb and Turner}
is the age of the flat universe and
$H_{\mathrm{c}}^{-1}/t_{\mathrm{a}} = 1.5$.

For the star collapse with the solar mass density
$\rho_{\odot} = 1.4 \,\mathrm{g}\,\mathrm{cm}^{-3}$,
we find $H_{\mathrm{s}}/H_{\mathrm{c}} = 
3.9 \times 10^{14}$,
where we have used
$3H_{\mathrm{s}}^2/\kappa^2 = \rho_{\odot}$ 
and the critical density 
$\rho_{\mathrm{crit}} \equiv
3H_{\mathrm{c}}^2/\kappa^2 = 9.2 \times 10^{-28} \, 
\mathrm{g}\,\mathrm{cm}^{-3}$~\cite{Kolb and Turner}.
In this case, from Eq.~(\ref{eq:B-3B-8}) we obtain
$t_{\mathrm{s}}/t_{\mathrm{a}} = 2.9 \times 10^{14} \e^{-1.5 \times 10^{29}}
\mathcal{I}$. If $\mathcal{I} \sim O(1)$,
$t_{\mathrm{s}}/t_{\mathrm{a}} \ll 1$, i.e., $t_{\mathrm{s}}$ is
much smaller than the age of the universe.
The reason why the time scale for the appearance of the
curvature singularity in the exponential model is so small
originates from the exponential factor $\e^{-R/R_{\mathrm{c}}}$.
%%%%%
We remark that 
if we apply the above analysis to 
cosmological circumstances, in which $H_{\mathrm{s}} \sim H_{\mathrm{c}}$, 
it follows from Eq.~(\ref{eq:B-3B-8}) 
that $t_{\mathrm{s}}/t_{\mathrm{a}} \sim 0.27 \mathcal{I}$.
As a consequnce, if $\mathcal{I} \sim O(1)$,
the time scale for the appearance of the
curvature singularity in the universe
is as long as the age of the universe.
%%%%%

%%%%%%%%%%%%%%%%%%%
%%%  Sec. IV
%%%%%%%%%%%%%%%%%%%
\section{Conclusion}

In the present paper, we have studied
a curvature singularity appearing 
in the star collapse for some models of $F(R)$ gravity. 
We have explored the mechanics to generate 
the curvature singularity. 
Furthermore, 
we have 
explicitly illustrated that the higher derivative term 
$R^\alpha$ ($1 < \alpha \leq 2$) could cure the curvature singularity 
and viable $F(R)$ gravity models 
could become free of such a singularity. 
It is quite remarkable that same scenario works to cure star singularity 
as well as finite-time future singularity. 
In addition, 
we have discussed the realization process of the curvature singularity 
in the viable $F(R)$ gravity model 
and estimated the time scale of its appearance 
in exponential gravity. 
It has been shown that in case of the star collapse, 
the time scale is much shorter than the age of the universe,
whereas in cosmological circumstances, 
it is as long as the cosmological time.

%%%%%%%%%%%%%%%%%%%%%%%%
%%%  Acknowledgments
%%%%%%%%%%%%%%%%%%%%%%%%
\section*{Acknowledgments}

The work is supported in part
by Global COE Program
of Nagoya University (G07) provided by the Ministry of Education, Culture,
Sports, Science \& Technology
(S.N.);
the JSPS Grant-in-Aid for Scientific Research (S) \# 22224003 (S.N.);
%%%
and
MEC (Spain) project FIS2006-02842 and AGAUR (Catalonia) 2009SGR-994
(S.D.O.).
%%%

\end{document}